\documentclass{elsart}
\journal{Phys. Lett. B}
\date{2008 August 27}
\usepackage{graphicx}
\usepackage{cite}
\usepackage{txfonts}

\begin{document}

\begin{frontmatter}

\begin{flushright}
{\small {CERN-PH-TH/2007-137, MPP-2007-112, {\tt arXiv:0708.2889v3 [astro-ph]}}}
\end{flushright}
\vspace*{-.2cm}

\title{Probing quantum gravity using photons from a flare
       of the active galactic nucleus Markarian 501 observed by
       the MAGIC telescope}

\collab{MAGIC Collaboration}
\author[a]{J.~Albert},
\author[b]{E.~Aliu},
\author[c]{H.~Anderhub},
\author[d]{L.~A.~Antonelli},
\author[e]{P.~Antoranz},
\author[f]{M.~Backes},
\author[g]{C.~Baixeras},
\author[e]{J.~A.~Barrio},
\author[h]{H.~Bartko},
\author[i]{D.~Bastieri},
% ...
%\vspace*{-0.7cm}
%\collab{(The MAGIC Collaboration)}
\collab{and}
\author[z]{John Ellis},
\author[aa]{N.~E. Mavromatos},
\author[ab,ac,ad]{D.~V. Nanopoulos},
\author[c,z]{A.~S. Sakharov},
\author[z,ae]{E.~K.~G. Sarkisyan\thanksref{sark}}

\address[a]{Universit\"at W\"urzburg, D-97074 W\"urzburg, Germany}
\address[b]{IFAE, Edifici Cn., Campus UAB, E-08193 Bellaterra, Spain}
\address[c]{ETH Zurich, CH-8093 Switzerland}
\address[d]{INAF National Institute for Astrophysics, I-00136 Rome, Italy}
\address[e]{Universidad Complutense, E-28040 Madrid, Spain}
\address[f]{Technische Universit\"at Dortmund, D-44221 Dortmund, Germany}
\address[g]{Universitat Aut\`onoma de Barcelona, E-08193 Bellaterra, Spain}
\address[h]{Max-Planck-Institut f\"ur Physik, D-80805 M\"unchen, Germany}
\address[i]{Universit\`a di Padova and INFN, I-35131 Padova, Italy}
% ...
\address[z]{Physics Department, CERN, CH-1211 Geneva 23, Switzerland}
\address[aa]{King's College London, Department of Physics, Strand, London WC2R 2LS, UK}
\address[ab]{Texas A\&M University, College Station, TX-77843, USA}
\address[ac]{HARC, Woodlands, TX-77381, USA}
\address[ad]{Academy of Athens, Division of Natural Sciences, GR-10679 Athens, Greece}
\address[ae]{Physics Department, Universiteit Antwerpen, B-2610 Wilrijk, Belgium}

\corauth[cor]{Corresponding author.}
\thanks[otte]{Now at: University of California, Santa Cruz, CA-95064, USA}
\thanks[sark]{Now at: University of Texas, Arlington, TX-76019, USA}
\thanks[pann]{Deceased.}

\date{Received 18 August 2008; Accepted 27 August 2008}

\begin{abstract}
We analyze the timing of photons observed by the MAGIC telescope during a flare
of the active galactic nucleus Mkn 501 for a possible correlation with energy,
as suggested by some models of quantum gravity (QG), which predict a vacuum
refractive index $\simeq 1 + (E/M_{\rm QGn})^n$, $n = 1,2$. Parametrizing the
delay between $\gamma$-rays of different energies as $\Delta t =\pm\tau_l E$ or
$\Delta t =\pm\tau_q E^2$, we find $\tau_l=(0.030\pm0.012)$~s/GeV at the
2.5-$\sigma$ level, and $\tau_q=(3.71\pm2.57)\times10^{-6}$~s/GeV$^2$,
respectively. We use these results to establish lower limits $M_{\rm QG1} >
0.21 \times 10^{18}$~GeV and $M_{\rm QG2} > 0.26 \times 10^{11}$~GeV at the
95\% C.L. Monte Carlo studies confirm the MAGIC sensitivity to propagation
effects at these levels. Thermal plasma effects in the source are negligible,
but we cannot exclude the importance of some other source effect.
\end{abstract}


\end{frontmatter}

\section{Introduction}
It is widely speculated that space-time is a dynamical medium, subject to
quantum-gravitational (QG) effects that cause space-time to fluctuate on the
Planck time and distance scales~\cite{amellis,foam,newstring,gambini,alfaro,kostel,amecam,myers}, 
for reviews see~\cite{review}. It has also been suggested that this `foaming' of
space-time might be reflected in modifications of the propagation of energetic
particles, namely dispersive effects due to a non-trivial refractive index
induced by the QG fluctuations in the space-time foam. There are microscopic
string-inspired models~\cite{amellis,foam,newstring} that predict only
subluminal refraction, and only for photons~\cite{crab}, suppressed either
linearly or quadratically by some QG mass scale:
\begin{equation}
\label{linear}
\frac{\Delta c}{c}=-\frac{E}{M_{\rm QG1}}~, \; {\rm or} \;
\frac{\Delta c}{c}=-\frac{E^2}{M^2_{\rm QG2}}.
\end{equation}
One might guess that the scale
$M_{\rm QG1}$ or $M_{\rm QG2}$ would be related to
${\hat M}_{\rm P}$, where ${\hat M}_{\rm P} =
2.4 \times 10^{18}$~GeV is the reduced Planck mass, but smaller
values might be possible in some string theories~\cite{foam,newstring}, or
models with large extra dimensions~\cite{merab}. Superluminal
modes and birefringence effects are also allowed in some other
models~\cite{gambini,alfaro,kostel,amecam,myers}.

A favored way to search for such a non-trivial dispersion relation is to
compare the arrival times of photons of different energies arriving on Earth
from pulses of distant astrophysical sources~\cite{amellis,mavrik}.
The greatest sensitivities may be expected from sources with short pulses, at
large distances or redshifts $z$, of photons observed over a large range of
energies. In the past, studies have been made of emissions from
pulsars~\cite{pulsar}, $\gamma$-ray bursts
(GRBs)~\cite{amellis,mavrik,wavegrb,robust,robust2,piranishvilo,merab} and
active galactic nuclei (AGNs)~\cite{mkr421,BLM}.
In particular, a combined analysis of many GRBs at different redshifts made
possible some separation between energy- and source-dependent effects, and
yielded a robust lower limit $M_{\rm QG1} > 0.9 \times
10^{16}$~GeV~\cite{robust}. Astrophysical sources that produce very high
energy photons in the TeV range or higher could improve significantly the
sensitivity to Lorentz violation, if one could distinguish source and
propagation effects. Flaring AGNs are celestial objects with the desired
properties, and a pioneering study of a flare of the AGN Mkn~421 yielded a
sensitivity to $M_{\rm QG1} \sim 4 \times
10^{16}$~GeV~\cite{mkr421}.\footnote{Stronger limits hold in models predicting
birefringence~\cite{uvgal,uvgrb}, but these do not apply to stringy models of
QG-induced vacuum dispersion~\cite{foam,newstring}, in which birefringence is
absent.}
 
In this Letter we analyze two flares of Mkn~501 ($z = 0.034$) observed by the
Major Atmospheric Gamma-ray Imaging Cerenkov (MAGIC) telescope~\cite{magic}
between May and July 2005. After applying standard quality checks, data
covering a total observation time of 31.6~h spread over 24 nights survived and
were analysed~\cite{magicobs}. The data were taken at zenith angles of
10$^\circ-$30$^\circ$, resulting in an energy threshold (defined as the peak of
the differential event-rate spectrum after cuts) of $\approx 150$ GeV.
The air-shower events were subjected to the standard MAGIC analysis
\cite{magican}, which rejects about 99\% of hadronic background events, while
retaining 50$-$60\% of the $\gamma$-ray induced showers. The $\gamma$-ray
energies are, in a first approximation, proportional to the total amount of
light recorded in the shower images; corrections are applied according to
further image parameters \cite{hillaspar} obtained from the analysis. The
achieved energy resolution is $\sim 25$\% over the range 170~GeV to 10~TeV. The
arrival time of each event is obtained with sub-ms
precision.

During the observations, variations in the $\gamma$-ray flux by an order of
magnitude were observed, with the maximum integrated flux above $\approx$~150
GeV exceeding $(11.0\pm0.3) \times 10^{-10}$ cm$^{-2}$s$^{-1}$.
In the two nights with the highest flux, high-intensity outbursts of short
duration (flares) were recorded with characteristic rise and fall times of
1$-$2 minutes. While the flare of July~9 was clearly visible over the full
energy range 0.15$-$10~TeV and reached a peak flux more than a factor two
higher than before and after the flare, that seen on June~30 was concentrated
at low energies (0.25$-$1~TeV) and less significant. In the analysis below, we
applied cuts on the image parameter $ALPHA$ \cite{hillaspar}, describing the
gamma shower arrival direction: $|ALPHA| <10^\circ$, and on energy: $E_\gamma
>150$~GeV.

\section{Timing analysis}
The spectral time properties of the most intense portions of the flares were
quantified in~\cite{magicobs} using four different energy bands with
boundaries at 0.15, 0.25, 0.6 and 1.2~TeV, the fourth band extending to
infinite energies. In the June~30 flare a signal above a uniform background
appeared only in the energy band of 0.25$-$0.6 TeV, which did not permit any
conclusion on the time-spectral properties of the signal. For the flare of
July~9, a time lag of about 4 minutes was found for the maximum of the time
profile envelope for photons in the 1.2$-$10~TeV energy band relative to those
in the range 0.25$-$0.6~TeV. The difference between the mean energies in these
two bands is $\approx$~2 TeV, which would lead to a naive estimate of a time
delay of about 0.12 s for each GeV of energy difference.
However, this approach is too simplistic, since the energy range covered by the
1.2$-$10~TeV band is much larger than the energy difference between the two
bands, so the binned estimator used in~\cite{magicobs} is inadequate for
constraining possible linear or quadratic energy dependences. In view of this
and the limited number of photons, we improve here on the binned estimator by
analyzing the complete information encoded in the time-energy distribution of
{\it individual} photons in the flare, with the aim of probing possible
systematic energy-dependent time lags induced by QG vacuum refraction during
photon propagation to the Earth, or intrinsic to the source.
\begin{figure}
\begin{center}
\begin{minipage}{97mm}
\includegraphics[width=\linewidth]{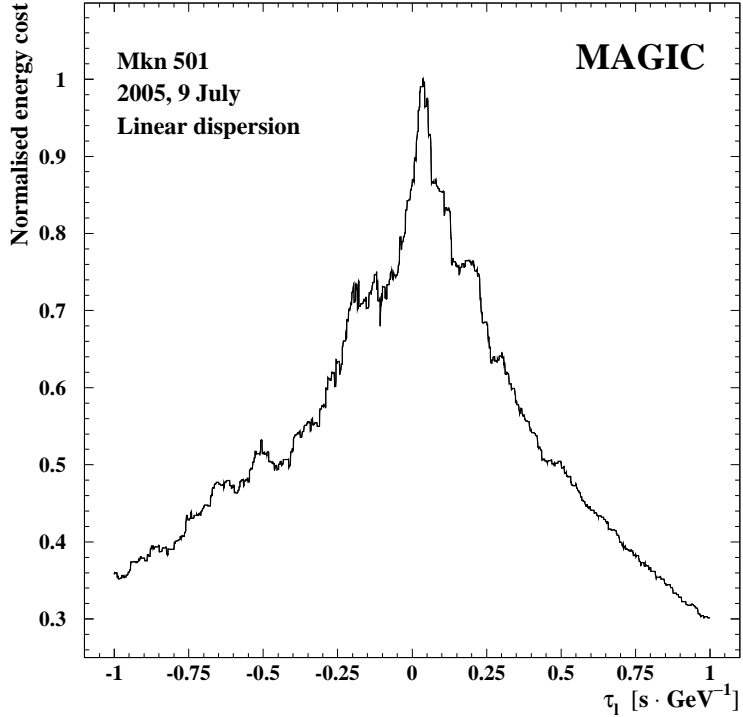}
\caption{The ECF from one realization of the MAGIC measurements with
photon energies smeared by Monte Carlo, for the case of a vacuum refractive
index that is linear in the photon energy.}
\label{costeqg}
\end{minipage}
\end{center}
\end{figure}
The true shape of the time profile at the source is not known, so we choose the
following analysis strategy. In general, the short pulse structure of any
flare would be blurred by an energy-dependent effect on photon propagation.
Conversely, one may correct for the effects of any given parametric model of
photon dispersion, e.g., the linear or quadratic vacuum refractive index, by
applying to each photon the appropriate time shift~\cite{robust} corresponding
to its propagation in a spatially-flat expanding universe:
$\Delta t(E)=H_0^{-1}(E/M_{\rm QG1})\int\limits_0^z(1+z)h^{-1}(z)\mathrm{d}z$
or similarly for the quadratic case, where $H_0$ is the Hubble expansion rate
and $h(z) = \sqrt{\Omega_{\Lambda} + \Omega_M (1 + z)^3}$. If the correct
energy-dependent QG shift is applied, the short pulse structure
of the emission profile is restored.

We implement this analysis strategy in two ways. In one analysis, we consider
the most active part of the flare, that is distinguished clearly from the
uniform background, and the QG shift is varied so as to maximize the total
energy in this part. In the other analysis we use the shape of the flare as
extracted from untransformed low-energy data.
\begin{figure}
\begin{center}
\begin{minipage}{97mm}
\includegraphics[width=\linewidth]{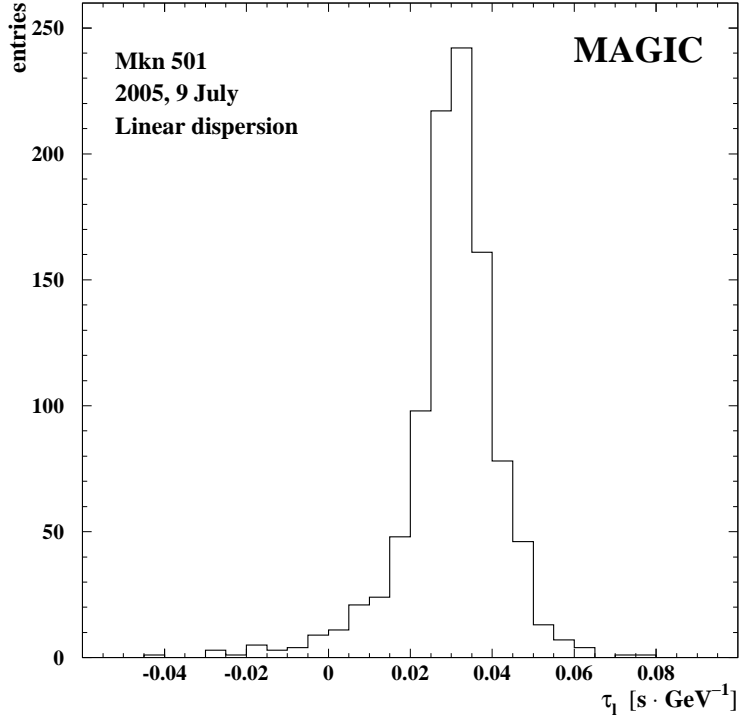} 
\caption{The $\tau_l$ distribution from fits to the ECFs of 1000 realizations
of the July~9 flare with photon energies smeared by Monte Carlo.}
\label{histas}
\end{minipage}
\end{center}
\end{figure}

\subsection{Energy cost function}
It is well known~\cite{jac} that a pulse of electromagnetic radiation
propagating through a linearly-dispersive medium, as postulated here, becomes
diluted so that its power (the energy per unit time) decreases.\footnote{The
applicability of classical electrodynamics for estimating the low-energy
behavior induced by space-time
foam~\cite{review,amellis,foam,gambini,alfaro,kostel,amecam,myers,merab} and
the corresponding pulse-broadening effect have been discussed elsewhere
(see~\cite{mavrik} for details and an explicit example). The dilution effects
for the linear or quadratic cases may easily be obtained as described in
\cite[Sect 7.9]{jac} by applying the dispersion laws $\omega
(k)=k[1-k/(2M_\mathrm{QG1})]$ or $\omega (k)=k[1-k^2/(3M_\mathrm{QG2}^2)]$.}
Any transformation of a signal to reproduce the undispersed signal tends to
recover the original power of the pulse. If the parameter $M_{\rm QGn}$
is chosen correctly, the power of the recovered pulse is maximized.

We implement this observation as follows. First, we choose a time interval
$(t_1;t_2)$ containing the most active part of the flare, as determined using a
Kolmogorov-Smirnov (KS) statistic~\cite{ks}. The KS statistic is calculated
from the difference between the cumulative distribution function estimated from
the unbinned data and that of a uniform distribution. The interval $(t_1;t_2)$
covers the time range where the value of the KS difference varies from its
maximum over the whole time support of the signal down to a negligible value.
This procedure determines the proper time-width $t_2-t_1$ of the most active
(transient) part of the flare.\footnote{The time interval chosen agrees very
well with the spread of a Gaussian fit to the profile of the binned data, as
well as the more complicated profile used in~\cite{magicobs}.}
Having chosen this window, we scan over the whole support the time-distribution
of all photons shifted by $\Delta t(E)$ and sum up the energies of photons in
the window. For convenience, we re-parametrize the time shift as 
$\Delta t = \pm \tau_l E$ or $ \Delta t = \pm \tau_qE^2$ respectively,
with $\tau_l$ and $\tau_q$ in ${\rm s/GeV}$ and ${\rm s/GeV^2}$ units.
The transformation is repeated for many values of $\tau_l$ and $\tau_q$, chosen
so that the shifts $\Delta t$ match the precision of the arrival-time
measurements, and for each $\tau_l$ or $\tau_q$ the scan is performed and the
maximal summed energy in a window of width $t_2-t_1$ is obtained. The maximal
energies as a function of $\tau_l$ or $\tau_q$ define the `energy cost
function' (ECF). The position of the maximum of the ECF indicates the value of
$\tau_l$ or $\tau_q$ that best recovers the signal, in the sense of maximizing
its power.\footnote{Varying slightly the boundaries of the interval $(t_1;t_2)$
has a negligible effect on the position of the maximum. We take into account
the difference between the width at the Earth and at the source, also
negligible.} This procedure is applied to 1000 Monte Carlo (MC) data samples
generated by applying to the measured photon energies the (energy-dependent)
Gaussian measurement errors.

Fig.~\ref{costeqg} shows the ECF for one such energy-smeared MC sample. It
exhibits a clear maximum, whose position may be estimated by fitting it with a
Gaussian profile in the peak vicinity. Fig.~\ref{histas} shows the results of
such fits to the ECFs with $\tau_l$ for the 1000 energy-smeared realizations of
the July~9 flare. From this distribution we derive the value $\tau_l=(0.030\pm
0.012)$~s/GeV, where $M_{\rm QG1}=1.445\times10^{16}\,{\rm s}/\tau_l$, leading
to a lower limit $M_{\rm QG1} > 0.21 \times 10^{18}$~GeV at the 95\%
C.L.\footnote{ We propagate the large errors by using the $\pm 1$-$\sigma$
range of ${\hat M}_{\rm P}/M_{\rm QGn}$ to estimate the $\mp1$-$\sigma$ range
of $M_{\rm QGn}$.} The same procedure applied to the ECF obtained using
$\tau_q$ leads to $\tau_q=(3.71 \pm 2.57)\times 10^{-6}$~s/GeV$^2$, where
$M_{\rm QG2}=1.222\times10^8\,({\rm s}/\tau_q)^{1/2}$, corresponding to $M_{\rm
QG2} > 0.26 \times 10^{11}$~GeV at the 95\% C.L. While our results for the
June~30 flare have similar sensitivities and are compatible, they cannot be
used to strengthen our results, as this flare is not very
significant.

\subsection{Likelihood function}
We have confirmed this result using another technique to study the
energy-dependent delay signal in the data. It is motivated by the initial time
and energy-binned analysis performed in~\cite{magicobs}, which we used to check
that the light-curve is well described by a simple Gaussian profile,
superimposed on a time-independent background. We compute a likelihood function
${\cal L}$ based on the probability of a photon to be observed with energy $E$
and arrival time $t$, using variables describing the energy spectrum at the
source, the time distribution at emission obtained from the measured arrival
times of the photons assuming an adjustable energy-dependent propagation delay,
and the energy resolution of the detector, which is modelled as a
Gaussian~\cite{martinez}. To describe the photon energy at the source a simple
power law $\Gamma(E_{\rm s}) \sim E_{\rm s}^{-\beta}$ is taken, with $\beta =
2.7$ for the time-uniform part of the flare and $2.4$ for the flaring
part~\cite{magicobs}.
\begin{figure}
\begin{center}
\begin{minipage}{122mm}
\includegraphics[width=\linewidth]{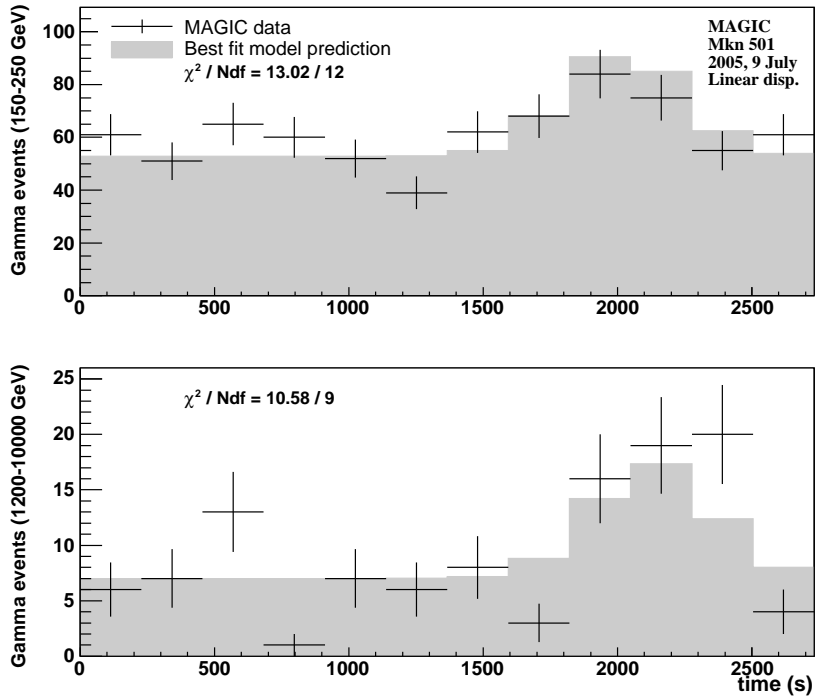}
\caption{Comparison of the MAGIC measured lightcurve at low and high
energies with the prediction given by the best set of parameters found using
the likelihood method, and binning the data and the likelihood function in the
same manner.}
\label{datafit}
\end{minipage}
\vspace*{.4cm}
\end{center}
\end{figure}

The likelihood function is fitted to the July~9 data minimizing $-\log{\cal L}$
as a function of four parameters: (i) the energy-dependent delay parameterized
in terms of ${\hat M_P}/M_{\rm QG1}$, (ii) the position of the intrinsic
maximum of the Gaussian flare, (iii) its width and (iv) the normalization of
the time-independent background component in arbitrary units. The best fit
yields ${\hat M}_{\rm P}/M_{\rm QG1} = 8.2^{+ 3.7}_{- 3.4}$, corresponding to 
$\tau_l = (0.048\pm0.021)$~s/GeV.
The shape of the function $\chi^2 \equiv -2 \log {\cal L} + {\rm const}$ around
the minimum in these variables is quite parabolic almost up to the 2-$\sigma$
level. In view of the correlations with these parameters, the sensitivity to
$\tau_l$ would be improved if they were known more precisely.

A similar procedure in the case of a quadratic dependency gives
$\tau_q = (4.60\pm5.46)\times10^{-6}$ s/GeV$^2$.

Fig.~\ref{datafit} shows that the ${\cal L}$ function gives a good overall fit
to the data: binning in time and energy both the data and the ${\cal L}$
function, we find $\chi^2$/NDF $\sim 1$.

\subsection{Crosscheck with Monte Carlo data}
To check the robustness of the ECF and likelihood analyses, we simulated
several MC test samples with two components: (a) a time-independent background
with the same energy spectrum as the measured data before the flare, and (b) a
superposed signal generated at the source with an energy spectrum similar to
that observed during the flare and an energy-independent Gaussian time
distribution, each with the same numbers of photons as in the measurement. We
then calculated the arrival times of all photons using various dispersion
models and parameters, taking into account the MAGIC energy resolution. For
each dispersion model and parameter, we generated 1000 incarnations, using
different random seeds. These samples were then analyzed blindly, and the
encoded effects were recovered successfully by the two estimators within the
expected uncertainties. In addition, the analysis techniques were applied to MC
samples with no energy-dependent dispersive signal encoded, and found no
effect, and both techniques also returned null results when applied to Mkn 501
data from outside a flare. These tests confirm the numerical sensitivities of
the analyses and the estimates of the uncertainties given above. For the
likelihood method, additional checks have been performed~\cite{martinez}
assuming different flare energy spectra and shapes, besides the Gaussian one
discussed here, which also fit reasonably well the binned data (c.f.
Fig.~\ref{datafit}).

\section{Conclusions}
The probability of the zero-delay assumption relative to the one obtained with
the ECF estimator is $P=0.026$.
The observed energy-dependent delay thus is a likely observation, but does not
constitute a statistically firm discovery.
The results of the two independent analyses of the July~9 flare of Mkn~501 are
quite consistent within the errors. Their results exhibit a delay between
$\gamma$-rays of different energies, $\tau_l=(0.030\pm0.012)$~s/GeV,
corresponding to a lower limit $M_{\rm QG1} > 0.21 \times 10^{18}$~GeV at the
95\% C.L. We also find a quadratic delay
$\tau_q=(3.71\pm2.57)\times10^{-6}$~s/GeV$^2$, and $M_{\rm QG2} > 0.26 \times
10^{11}$~GeV at the 95\% C.L., far beyond previous limits on a quadratic effect
in photon propagation~\cite{mkr421,wavegrb,merab}.
These numbers could turn into a real measurement of $M_{\rm QG1,2}$, if the 
emission mechanism at the source were understood and the observed delays were mainly
due to propagation.
We cannot exclude, however, the possibility that the delay we find, which is significant
beyond the 95\% C.L., is due to some energy-dependent effect at the
source.\footnote{Note that if the observed energy-dependent time shift is
explained by some source effect, the lower limit on $M_{\rm QG}$ would rise.}
However, we can exclude the possibility that the observed time delay may be due
to a conventional QED plasma refraction effect induced as photons propagate
through the source. This would induce~\cite{plasma} $\Delta t = D (\alpha^2
T^2/6q^2) \ln^2(qT/m_e^2)$, where $\alpha$ is the fine-structure constant, $q$
is the photon momentum, $T$ is the plasma temperature, $m_e$ is the mass of
electron, $D$ is the size of the plasma, and we use natural units: $c, \hbar
=1$. Plausible numbers such as $T \sim 10^{-2}$~MeV and $D \sim 10^9$~km
(for a review see \cite{hillas}) yield a negligible effect for $q \sim 1$~TeV.
Exclusion of other source effects, such as time evolution in the mean emitted
photon energy, might be possible with the observation of more flares, e.g., of
different AGNs at varying redshifts. Observations of a single flare cannot
distinguish the quantum-gravity scenarios considered here from modified
synchrotron-self-Compton mechanisms~\cite{magicobs,wagner08}. However, this
pioneering study demonstrates clearly the potential scientific value of an
analysis of multiple flares from different sources. The most promising
candidate for applying the analyses proposed here is the flare
from PKS 2155-304 detected recently by H.E.S.S. \cite{hess2155}. 
Unfortunately the occurrence of fast flares in AGNs is currently unpredictable,
and since no correlation has yet been established with observations in other
energy bands that could be used as a trigger signal, only serendipitous
detections are currently possible.

\section*{Acknowledgments}
The MAGIC Collaboration thanks the IAC for the excellent working conditions at
the Observatorio del Roque de los Muchachos in La Palma and gratefully
acknowledges the support of the German BMBF and MPG, the Italian INFN and the
Spanish CICYT. This work was also supported by ETH Research Grant TH 34/04 3
and the Polish MNiI Grant 1P03D01028. The work of J.E. and N.E.M. was
partially supported by the European Union through the Marie Curie Research and
Training Network UniverseNet MRTN-CT-2006-035863, and that of D.V.N. by DOE
grant DE-FG02-95ER40917.

\end{document}